\documentclass[11pt, a4paper]{article}
\usepackage{moriond,epsfig}
\usepackage{amssymb}
\usepackage{amsmath}

\def\be{\begin{equation}}
\def\ee{\end{equation}}
\def\bea{\begin{eqnarray}}
\def\eea{\end{eqnarray}}

\begin{document}
\vspace*{4cm}
\title{
INCLUSIVE HIGGS BOSON SEARCHES IN FOUR-LEPTON FINAL STATES \\ AT THE LHC
}

\author{ EVELYNE DELMEIRE \\ on behalf of the ATLAS and CMS Collaborations}
\address{ Universidad Aut\'onoma de Madrid \footnote{Now at the Universiteit Antwerpen.}}

\maketitle\abstracts{
The inclusive search for the Standard Model Higgs boson in four-lepton final states with
the ATLAS and CMS detectors at the LHC $\rm pp$ collider is presented.
The discussion focusses on the 
$ \rm H\rightarrow ZZ^{(*)}\rightarrow 4l+X$ 
decay mode for a Higgs boson in the mass range 
$ 120 ~\lesssim M_{\rm H} \lesssim ~600~{\rm GeV}/c^2$.
A prospective analysis is presented for the discovery potential
based on a detailled simulation of the detector response in the experimental
conditions of the first years of LHC running at low luminosity.  
An overview of the expected sensitivity in the
measurement of the Higgs boson properties is also given.} 
\noindent
\section{Introduction}
The Standard Model(SM) of electroweak interactions contains one
Higgs boson whose mass, $\rm M_{\rm H}$, is a free parameter of 
the model. The inclusive single production reaction $\rm p + p \rightarrow H + X $ 
followed by the decay $\rm H \rightarrow Z Z^{{(*)}} \rightarrow 
l^{+}l^{-}l'^{+}l'^{-}$(in short $\rm H \rightarrow 4l$)
is the cleanest("golden") decay mode for the discovery of the SM
Higgs boson at the LHC and can provide a sensitivity over a wide range 
of masses $\rm M_{\rm H}$ from 120 to 600 ${\rm GeV}/c^2$. 
There are three different final states which depend on the flavour of
the Z-boson decay leptons: $\rm H \rightarrow 4e$, $\rm H \rightarrow 4\mu$
and $\rm H \rightarrow 2e2\mu$.
Thanks to the relatively small background contamination, 
the $\rm H \rightarrow 4l$ also allows a precise measurement of the Higgs boson properties
(mass, width, spin, couplings, etc...).\\
\indent
The report summarizes the expected potential of ATLAS and CMS in SM Higgs boson searches
in the $\rm H \rightarrow 4l$ channel. 
For more details on the analyses described here, the reader is directed 
to the ATLAS~\cite{atlasphystdr} and  CMS~\cite{cmsphystdr2} Physics Technical Design Reports.
\section{Higgs boson signal and backgrounds}
At LHC energies, there are two dominant SM Higgs boson production processes:
gluon-gluon, $\rm gg \rightarrow H$, and weak boson fusion $\rm qq \rightarrow qqH$.
In the mass range $M_{\rm H}\lesssim 135 $ ${\rm GeV}/c^2$, the SM Higgs boson decays mainly into
$b\bar{b}$(BR $\approx$ 85\%) and $\tau^{+}\tau^{-}$(BR $\approx$ 8\%) pairs
but the search in the  $\rm H \rightarrow \gamma \gamma$ decay mode is priviledged 
despite it small branching ratio(BR $\approx$ 0.2\%) because of its clean experimental signature.
For $\rm M_{\rm H}\gtrsim 135 $ ${\rm GeV}/c^2$, the decay into $\rm H \rightarrow W^{+}W^{-}$
is dominant. The branching ratio for the $\rm H \rightarrow ZZ^{(*)}$ decay is 
sizable 
for $\rm M_{\rm H} \ge 115 ~{\rm GeV}/c^2$.\\
\indent
The $\rm H \rightarrow Z Z^{{(*)}} \rightarrow 4l$ signal event topology is
characterized by two pairs of oppositely charged and isolated 
same-flavour leptons coming from the same vertex with a di-lepton invariant mass
compatible with the Z-boson mass. The Higgs boson signal manifests itself as a narrow mass peak in
the reconstructed four-lepton invariant mass spectrum.
There are three main background sources to the $\rm H \rightarrow 4l$ signal.
The reducible $\rm Zb \bar{b}$ and $\rm t \bar{t}$ background
processes differ from the signal by the presence of
two non-isolated leptons inside b-jets with displaced vertices.
The irreducible $\rm ZZ^{(*)}$ background has kinematical properties 
which are very similar to that of the signal
except for the four-lepton invariant mass which shows a broad spectrum.
\section{Event selection}
All selections are optimised to have the highest significance for discovery
with emphasis on a realistic strategy for the control of experimental
errors and background systematics.\\
\indent
The on-line preselection consists of a logical OR of basic single and double
electron or muon triggers.
The off-line preselection starts with the search for events 
with at least four lepton candidates within the fiducial volume.
The aim is to reduce as much as possible the contamination
of background sources involving "fake" leptons from QCD jets
while preserving as much as possible the signal detection efficiency.
The $\rm Zb \bar{b}$ and $\rm t \bar{t}$ 
backgrounds have at least one non-isolated lepton-pair with often detectable displaced
vertices in contrast to the signal and $\rm ZZ^{(*)}$ background.
Therefore, the most descriminating preselection variables against these backgrounds
come from vertex constraints and isolation criteria relying on the measurement of
primary tracks in the tracker and/or the energy flow in the calorimeters.\\
\indent
The kinematical selection consists of cuts on the lepton transverse
momenta and the reconstructed di-lepton invariant mass spectra.
The first cut exploits the fact that b-decay leptons from the 
$\rm Zb \bar{b}$ and $\rm t \bar{t}$ backgrounds
have on average a softer $p_{T}$ spectrum than leptons from the Higgs boson signal or 
$\rm ZZ^{(*)}$ background.
The second requirement is powerful against all backgrounds.\\
\indent
The number of signal and background events is determined by a simple window sliding
in the reconstructed four-lepton invariant mass spectrum.
After the full selection, the reducible backgrounds are suppressed
well below the level of the $\rm ZZ^{(*)}$ contamination which remains the dominant and sole
remaining background. The typical rejection factors vary from
$2\times 10^{3}$ to $10^{4}$ for $\rm t \bar{t}$, from
$500$ to $10^{5}$ for $\rm Zb \bar{b}$
and from $20$ to $4$ for $\rm ZZ^{(*)}$, depending on the  $\rm M_{\rm H}$-hypothesis,
for a signal selection efficiency of 25-55 \%.
\begin{figure}
\begin{center}
\begin{tabular}{cc}
\hspace{-1.0cm}\psfig{figure=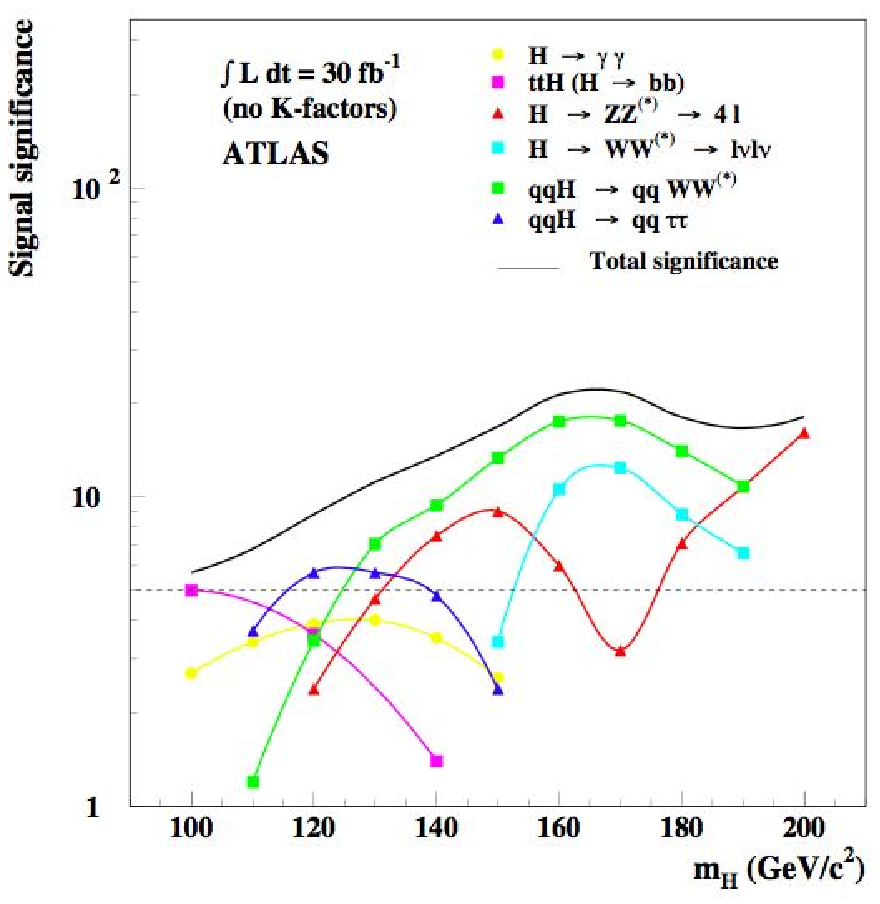,height=5.5cm} &\hspace{1cm}
\psfig{figure=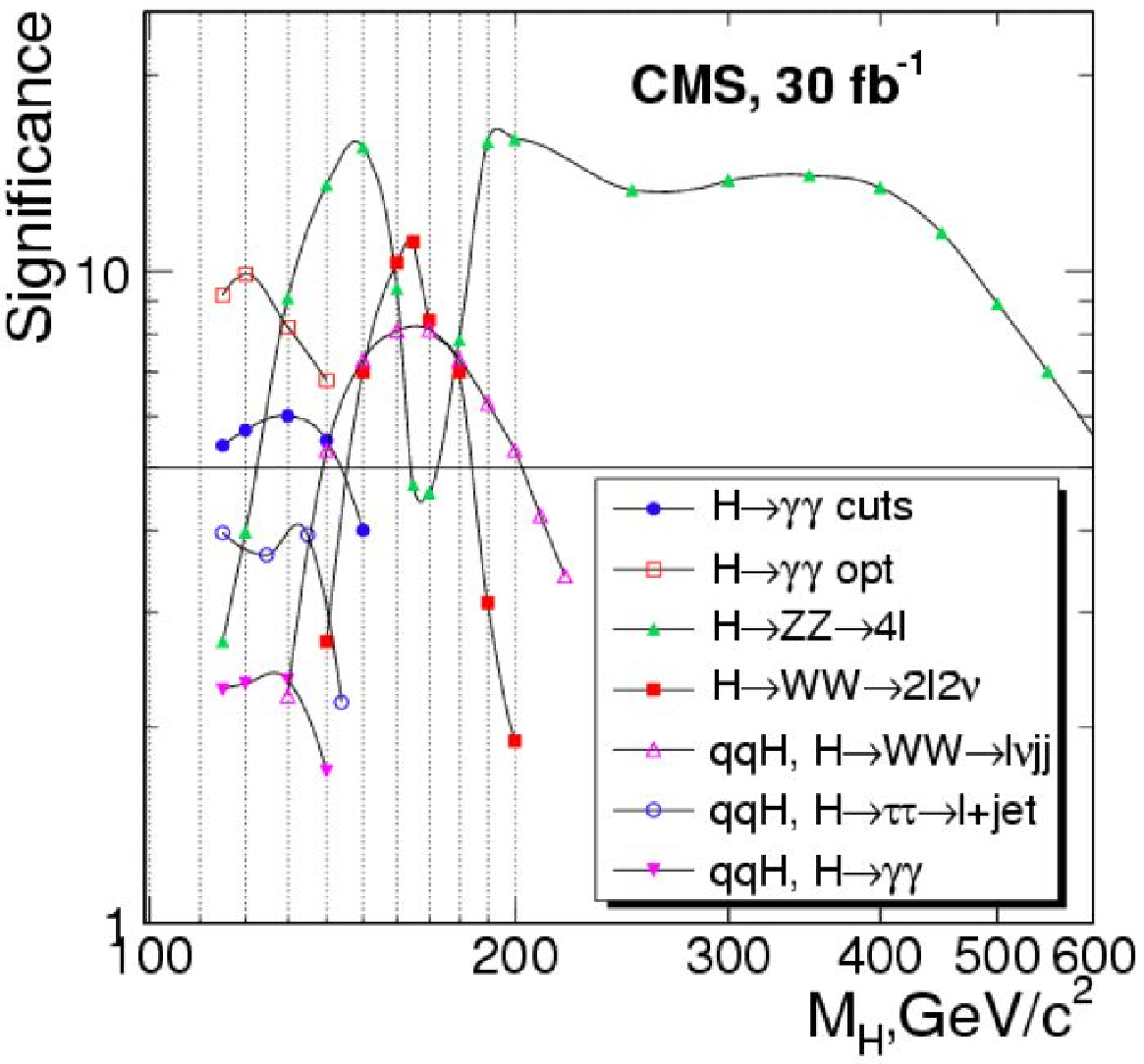,height=5.5cm}
\end{tabular}
\caption{The expected statistical significance for the Standard Model Higgs boson signal
as function of its mass 
for an integrated luminosity of 30 $\rm{fb}^{-1}$
for the ATLAS(Left) and CMS(Right) experiments.}
\label{fig:discovery}
\end{center}
\end{figure}
\section{Systematics}
The systematics on the signal significance are related to the knowledge
of the $\rm ZZ^{(*)}$ background rate in the signal region and 
the uncertainty on this knowledge.
Two approaches have been followed to estimate the background directly from the data: 
a normalisation to single Z, $\rm Z \rightarrow 2l$, data
and a normalisation to sidebands. 
Both approaches lead to a reduced sensitivity to theoretical 
and experimental uncertainties as well as a full cancellation of
the luminosity uncertainty. 
The theoretical uncertainty is of the order of 2 to 8\% for the normalisation
to  $\rm Z \rightarrow 2l$ and 0.5 to 4\% for the normalisation to sidebands.
The low statistics of $\rm ZZ^{(*)}$ events
could be a limiting factor for the sidebands method.\\
\indent
The overal strategy for controling the detector systematics is to estimate the
efficiency and the precision of the energy and momentum measurements from experimental
data. Single Z and single W processes have huge cross section at the LHC, and are
expected to lead to a significant reduction of the reconstruction uncertainties 
already after few $\rm{fb}^{-1}$.
\section{Discovery reach}
Figure~\ref{fig:discovery} shows the expected statistical significance for 
the SM Higgs boson signal as function of its mass 
for an integrated luminosity of 30 $\rm{fb}^{-1}$.
A 5$\sigma$-discovery is possible over a wide range of masses in the $\rm H \rightarrow 4l$
channel: 
$130 < \rm M_{\rm H}< 160 ~{\rm GeV}/c^2$ and $2 \rm m_{\rm Z}< \rm M_{\rm H}< 550 ~{\rm GeV}/c^2$.
The drop in sensitivity around $\rm M_{\rm H}\approx 180 ~{\rm GeV}/c^2$ 
will be filled by the complementary channel $\rm H \rightarrow WW^{(*)} \rightarrow 2l2\nu$ 
where less than 1 $\rm{fb}^{-1}$ is needed for 5$\sigma$-discovery. 
For $\rm M_{\rm H}< 130 ~{\rm GeV}/c^2$, the highest discovery potential
is obtained in the $\rm H \rightarrow \gamma\gamma$ decay mode.
\section{Measurement of the Higgs boson properties}
\begin{figure}
\begin{center}
\begin{tabular}{cc}
\hspace{-1.0cm}\psfig{figure=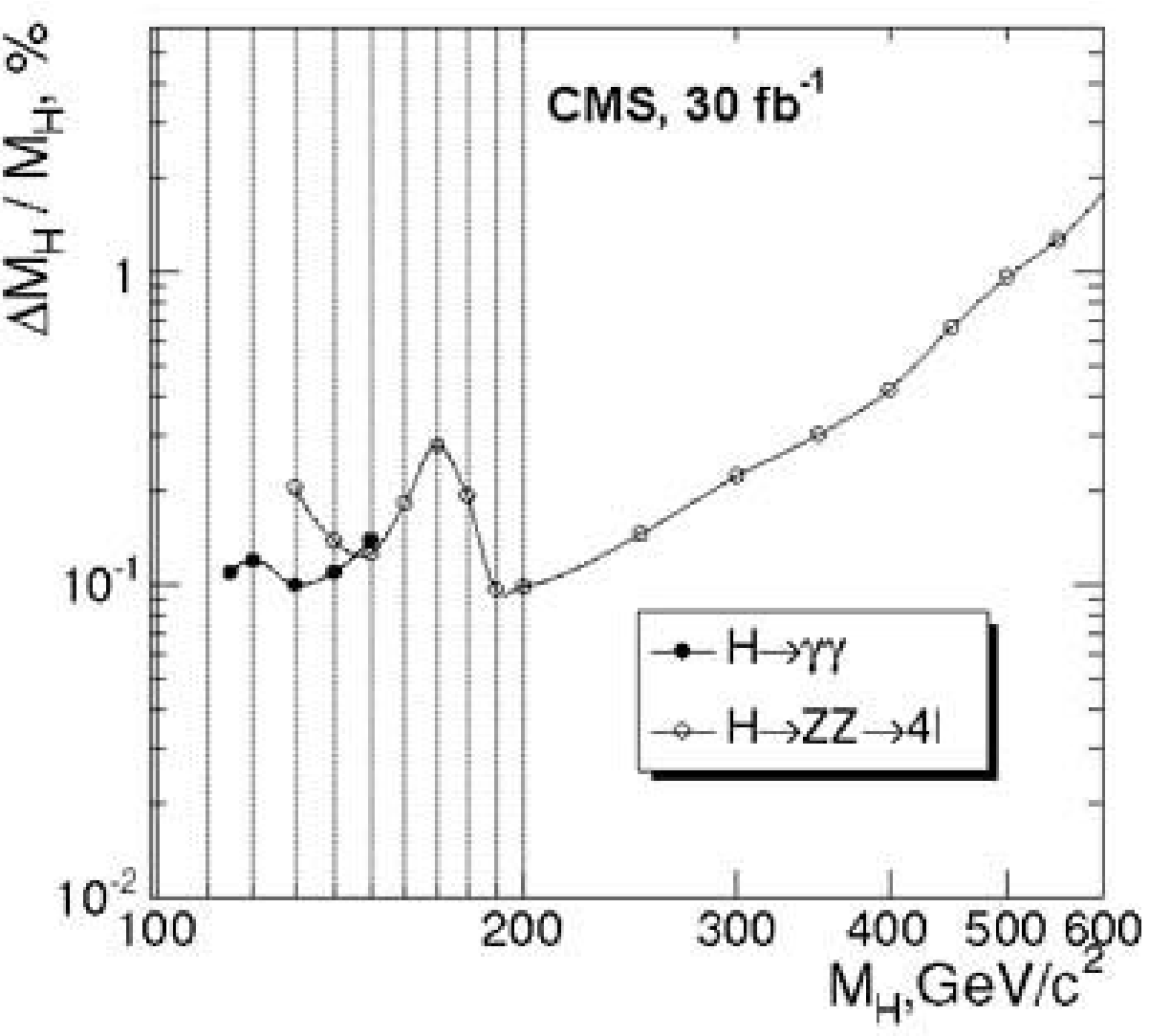,height=5cm} &\hspace{1cm}
\psfig{figure=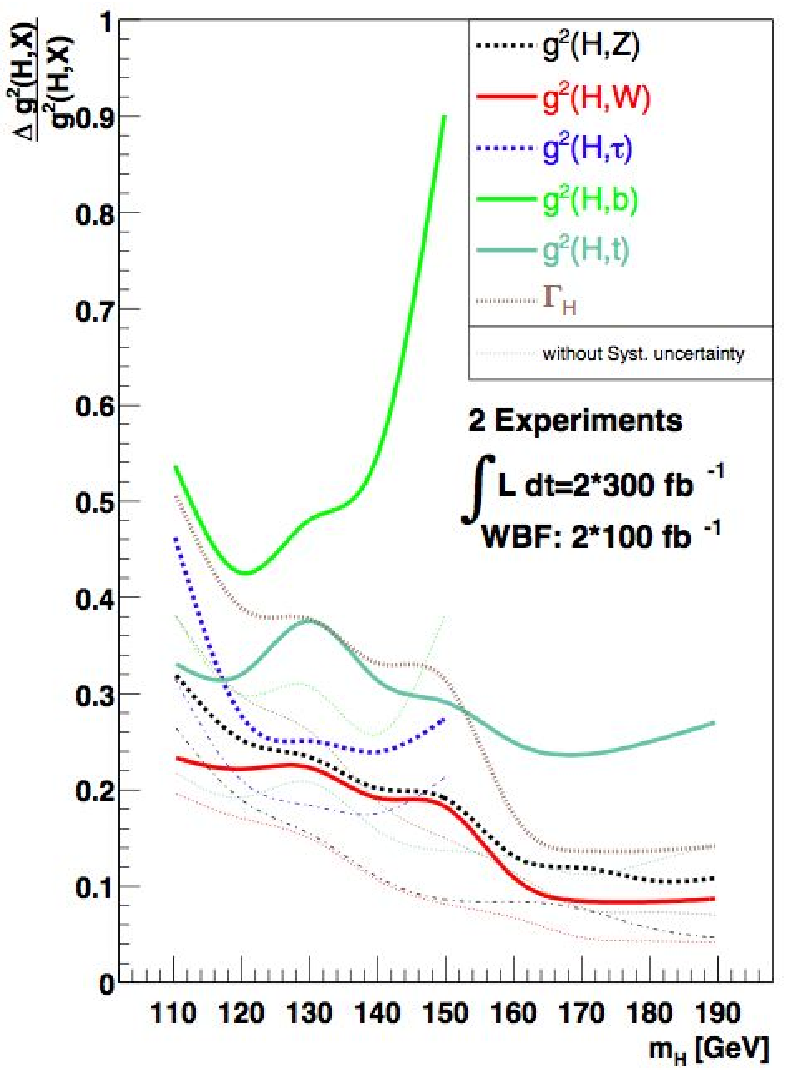,height=6.5cm}
\end{tabular}
\caption{
Left: The expected statistical precision of the Higgs boson mass measurement 
in CMS for an integrated luminosity of 30 $\rm{fb}^{-1}$. Right: 
The relative precision on the Higgs boson couplings assuming 300 $\rm{fb}^{-1}$ 
of data collected by both the ATLAS and the CMS experiments.
}
\label{fig:properties}
\end{center}
\end{figure}
The Higgs boson mass and width are obtained from a fit to the four-lepton invariant mass
spectrum. For an integrated luminosity of 30 $\rm{fb}^{-1}$, the expected statistical 
precision of the mass measurement is better than 1\% over a wide range of 
masses(Figure~\ref{fig:properties} Left).
The Higgs boson width measurement is only possible for Higgs boson masses beyond
200  ${\rm GeV}/c^2$ when the Higgs boson natural width starts to dominate the experimental resolution.
The expected precision on the width is smaller than 30\%.\\
\indent
The Higgs boson couplings to fermions and gauge bosons can be extracted from rate measurements in the 
different Higgs boson production and decay channels. Relative precision on the squared 
Higgs boson couplings, assuming 300 $\rm{fb}^{-1}$ of data collected by both the ATLAS and the CMS 
experiments, varies between 10\% and 40\% depending on the coupling, except for the Yukawa 
coupling to the b-quark which suffers from large uncertainties related to b-tagging and 
background normalisation~\cite{atlascouplings}.\\
\newpage
\indent
The $\rm H \rightarrow 4l$ channel is particularly suitable to measure the Higgs
boson spin and CP state because its small background contamination and the fact that the 
event kinematics can be completely reconstructed with good precision.
The angular correlations between the Z-boson decay products are used to extract the 
spin and CP state of the resonance.
A study based on ATLAS fast simulation shows that
with 100 $\rm{fb}^{-1}$ a
pseudo-scalar Higgs boson can be ruled out 
if $\rm M_{\rm H}> 200 ~{\rm GeV}/c^2$ and an
axial vector and vector  Higgs boson can be excluded
if $\rm M_{\rm H}> 230 ~{\rm GeV}/c^2$~\cite{atlasspincp}.
A recent analysis by the CMS experiment considers also
CP-violating spin-0 Higgs boson states via the introduction of a CP-mixing parameter
and determines the minimal enhancement or suppression in cross section
needed in  order to exclude the SM pseudo-scalar Higgs boson.
It is shown that the distinction between a scalar and a pseudo-scalar Higgs boson is 
already possible with 60 $\rm{fb}^{-1}$ integrated luminosity~\cite{cmsspincp}.
\section*{Acknowledgments}
The author would like to thank M.Aldaya, S.Baffioni, F.
Beaudette, C.Charlot, A.Drozdetskiy, D.Futyan, 
A.Nikitenko and I.Puljak for many interesting and helpful discussions

\end{document}